\documentstyle[here]{mn2e}

\def\gtsima{$\; \buildrel > \over \sim \;$}
\def\ltsima{$\; \buildrel < \over \sim \;$}
\def\gtrsim{\lower.5ex\hbox{\gtsima}}
\def\lesssim{\lower.5ex\hbox{\ltsima}}

\newcommand{\msun}{M$_{\odot}$}

\input{psfig}

\begin{document}

\title[Largest ring galaxy]{UGC 7069: The largest ring galaxy}
\author[Ghosh and Mapelli]
{Kajal K. Ghosh$^{1}$ and Michela Mapelli$^{2}$
\\
$^{1}$USRA/NSSTC/MSFC/NASA, 320 Sparkman Drive, Huntsville, AL35805, USA; {\tt kajal.k.ghosh@nasa.gov}\\
$^{2}$ Institute for Theoretical Physics, University of Z\"urich, Winterthurerstrasse 190, CH-8057, Z\"urich, Switzerland; {\tt mapelli@physik.unizh.ch}\\
}
\maketitle \vspace {7cm }

  \begin{abstract}
We find that UGC~7069 is the largest ring galaxy known to date. In this Letter, we present a multiwavelength study of this galaxy (combining radio, 2MASS, optical and ultraviolet data). The ring of UGC~7069, whose diameter measures $\sim{}115$ kpc, is also warped at its edges. The nucleus appears double-peaked and hosts a possible LINER. The ultraviolet data indicate a strong blue colour and suggest that UGC~7069 is a starburst galaxy. We also present $N$-body simulation results, which indicate that galaxy collisions can produce such huge rings. Large inclination angles between the target and the intruder galaxy may account for the formation of warped rings. Multiwavelength observations are highly essential to constrain our simulation results, which will address the formation and evolution of such a rare galaxy. 
\end{abstract}

%

\section{Introduction}

Ring galaxies are among the most fascinating objects in the sky. Their bright ring is often the site of intense star formation, as revealed by the H$\alpha{}$ and ultraviolet (UV) data. 
Such rings, at least that appear knotty and irregular, are thought to have formed through galaxy interactions (Lynds \& Toomre 1976; Theys \& Spiegel 1976; Appleton \& Struck-Marcell 1987a, 1987b; Hernquist \& Weil 1993; Mihos \& Hernquist 1994;  Appleton \& Struck-Marcell 1996; Struck 1997; Horellou \& Combes 2001; Mapelli et al. 2008a, 2008b). Thus, the dynamical evolution of ring galaxies is also particularly interesting. Simulations show that the ring is short-lived ($\lesssim{}500$ Myr) and that ring galaxies rapidly evolve into different objects, such as giant low surface brightness galaxies (e.g. Malin 1, Mapelli et al. 2008b) or recycled dwarf galaxies (e.g. NGC~5291; Bournaud et al. 2007).

The most famous ring galaxy, the Cartwheel galaxy, was also the largest one known so far (Struck et al. 1996; Borne et al. 1996, 1997). However, we find that the ring galaxy UGC~7069 has a physical diameter around $\sim$120~kpc, approximately twice as large as that of Cartwheel.
UGC~7069 was detected as an emission-line galaxy during the Kitt Peak National Observatory International Spectroscopic Survey (KISSR 1205, Salzer et al. 2005) and recently it has been observed during the Sloan Digital Sky Survey (SDSS) (SDSS J120457.92+430858.5, redshift = 0.05205).
It is a luminous member of the Two-Micron All-Sky Survey (2MASS) Flat Galaxy Catalogue and it was not covered in the IRAS survey.
 Its huge ring appears to be warped at the edges, a peculiarity which has never been observed in ring galaxies.  However, we stress that other ring galaxies may be warped (e.g. Cartwheel, Borne et al. 1997), but this feature can be clearly observed only when the inclination angle is sufficiently large. In addition, we have found two nuclei at the centre of  UGC 7069. Here, we present all the available multiwavelength data of UGC 7069, which show  that this is a starburst galaxy with high  star formation rate (SFR).
We also show $N$-body/smoothed particle hydrodynamics (SPH) models of this galaxy, which suggest that the warped ring formed after a galaxy interaction with very large inclination angle. 
Observations, data analysis and results are presented in Section~2. Details of simulations are presented in Section~3. Discussion and conclusions are described in Section~4. 
We adopt $H{_0}$ = 73~km~s$^{-1}$~Mpc$^{-1}$, $\Omega_{M}$ = 0.24, $\Omega_{\Lambda}$ = 0.76 (Spergel et al. 2007). The cosmology corrected  luminosity distance, distance modulus ($m-M$), and angular-size scale are 226 Mpc, 36.77 mag and 989~pc~arcsec$^{-1}$, respectively.

\section{Observations, data analysis and results}
\begin{figure}
\centerline{\psfig{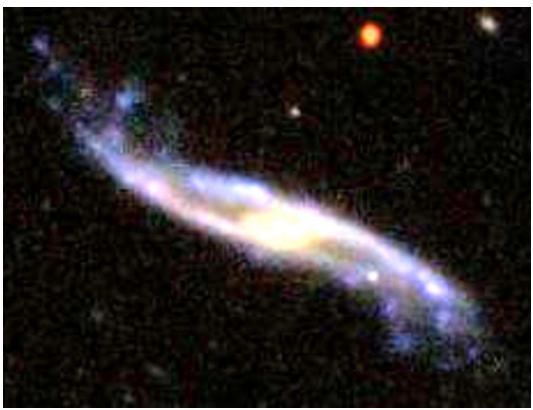}}
\caption{SDSS composite image of UGC~7069. The original image has been enhanced to show the faint features present at the ends of this galaxy. Its central region shows the presence of a nuclear disc, which is surrounded by a huge star forming ring. Large number of star-forming complexes are present on this ring.}
\end{figure}

\begin{figure*}
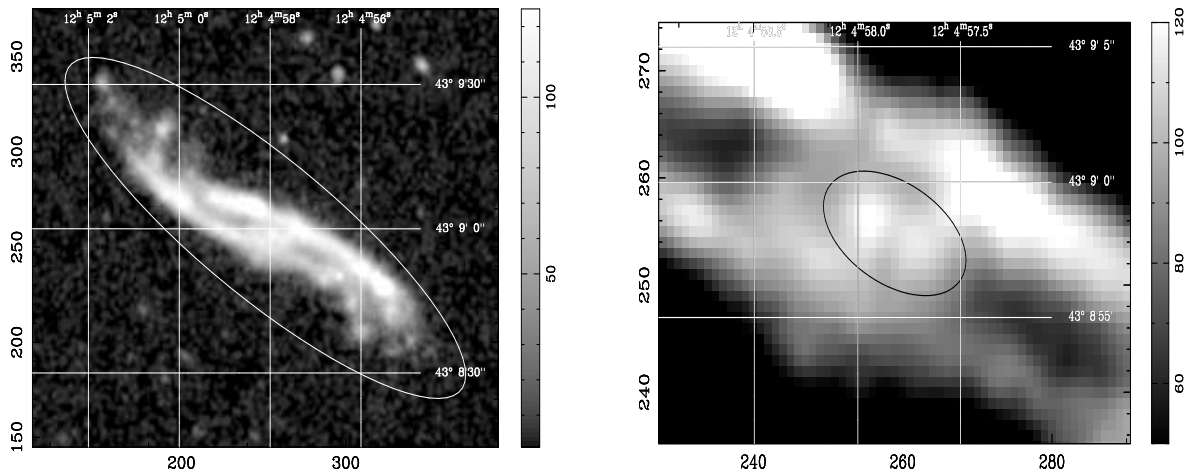

\centerline{\psfig{figure=new_large_galaxy_B.ps,angle=-90,height=6.15cm,width=7.25cm}
\hskip 1cm
            \psfig{figure=new_large_galaxy_B_zoom.ps,angle=-90,height=6.15cm,width=7.25cm}}
\caption{Left-hand panel (Fig. 2a): the same as Fig. 1 but for the $g$ band. The major- and minor-axis diameters of the $D_{23}$ isophote are 116".0 and 34".0, respectively. Right-hand panel (Fig. 2b) shows the nuclear region of Fig. 2a. Small ellipse at the centre is drawn around two objects, which may be the two nuclei of UGC~7069.}
\end{figure*}

According to the NASA/IPAC Extragalactic Database (NED), UGC~7069 is a SBcd starburst galaxy at a redshift of 0.052 with major- and minor-axis diameters $\sim$108".0 and $\sim$13".8, respectively. We have retrieved the photometric and spectroscopic data from the SDSS database. Photometric data were analyzed using our in-house software package, $LEXTRCT$ (Tennant 2006). The spectroscopic data were analyzed using the IRAF software package. The SDSS-composite image of UGC~7069 is displayed in Fig.~1. It can be seen from this figure that a huge extended disc (star-forming ring) surrounds the nuclear-disc ($\sim$7~kpc in diameter). In order to determine the size of the whole galaxy, we displayed the SDSS/$g$-band image of UGC~7069 in Fig. 2a. $D_{23}$ ellipse with major- and minor-axis diameters as 116".0 and 34".0, respectively, that was estimated from this image, is also shown in this figure. This suggests that the major-axis diameter of UGC~7069 is around 115~kpc. Deep images of UGC~7069 will reveal the true size of this galaxy. In Fig.~2b, we show the zoomed and contrasted image of the nuclear region of UGC~7069. A small ellipse is shown around the centre of this galaxy and it can be seen that there are two objects within the ellipse. Most likely, these two objects are the possible nuclei of the galaxy. If this interpretation is correct, then the existence of these two nuclei poses great challenge to the formation theory of such a giant system (see Section~3).

The nature of the two nuclei can be determined from the spectrum of the nuclear region of this galaxy. Optical fibre of the SDSS spectrograph has covered both the nuclei of UGC~7069.  Weak emissions are present only in a couple of lines around H$\alpha$. It may be possible that the underlying Balmer absorption from A-type stars present in the nuclear region of this galaxy weakens the H$\alpha$ emission. Spectrum of UGC~7069 in the H$\alpha$ region is shown in Fig. 3, which clearly displays that only [NII 6548\AA], H$\alpha$ (6563 \AA), [NII 6583\AA], [SII 6717 \AA] and [SII 6731 \AA] lines are in emission. Equivalent widths of these emission lines are -3.7, -10.9, -7.9,  -4.1 and -2.1~\AA, respectively. The intensity ratio of 
[NII 6583\AA] and  H$\alpha$ (6563 \AA) emission lines is around 0.72$\pm$0.1. This suggests that the nuclei may be low-ionization nuclear emission-line regions (LINERs), because this ratio is less than 0.6 for HII regions (Veilleux \& Osterbrock 1987; Ho, Filippenko \& Sargent 1997a). However, the full width at half intensity maximum of the  H$\alpha$ line is only $\sim$580~km~s$^{-1}$. This is indicative of the absence of pure active galactic nuclei (AGNs) in these LINERs (Ho, Filippenko \& Sargent 1997b; Maoz et al. 1998). Salzer et al. (2005) obtained a long-slit spectrum of this galaxy. From this spectrum, they detected  H$\beta$, [OIII~5007\AA], [NII 6548\AA], H$\alpha$ (6563 \AA), [NII 6583\AA], [SII 6717 \AA] and [SII 6731 \AA] emission lines. Based on the observed line-ratios of these emission lines, they concluded that  UGC~7069 is a starburst galaxy (Salzer et al. 2005). A comparison between the nuclear spectrum and the  long-slit spectrum of this galaxy clearly suggests that the contribution of the ring to star formation is dominant. This is consistent with the observations of other ring galaxies (e.g. Cartwheel, Borne et al. 1997). Two-dimensional spectra of this galaxy are needed to accurately determine the global SFR, metallicities, velocity maps, etc. All these measured parameters will be very useful to constrain the $N$-body/SPH model of UGC~7069, which is described in the following section.

\begin{figure}
\centerline{\psfig{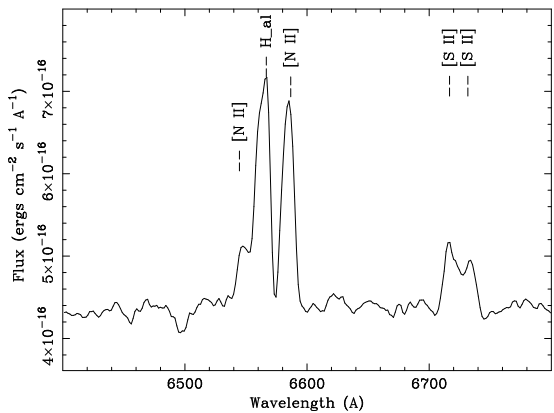}}
\caption{SDSS spectrum  of the nuclear region of UGC~7069. This  has been corrected for the Galactic extinction and the wavelength scale has been shifted to the source frame (redshift = 0.05205). Here we display the spectrum only in the H$\alpha$ region, which shows the emission lines of [NII 6548\AA], H$\alpha$ (6563 \AA), [NII 6583\AA], [SII 6717 \AA] and [SII 6731 \AA].}
\end{figure}

\begin{figure*}
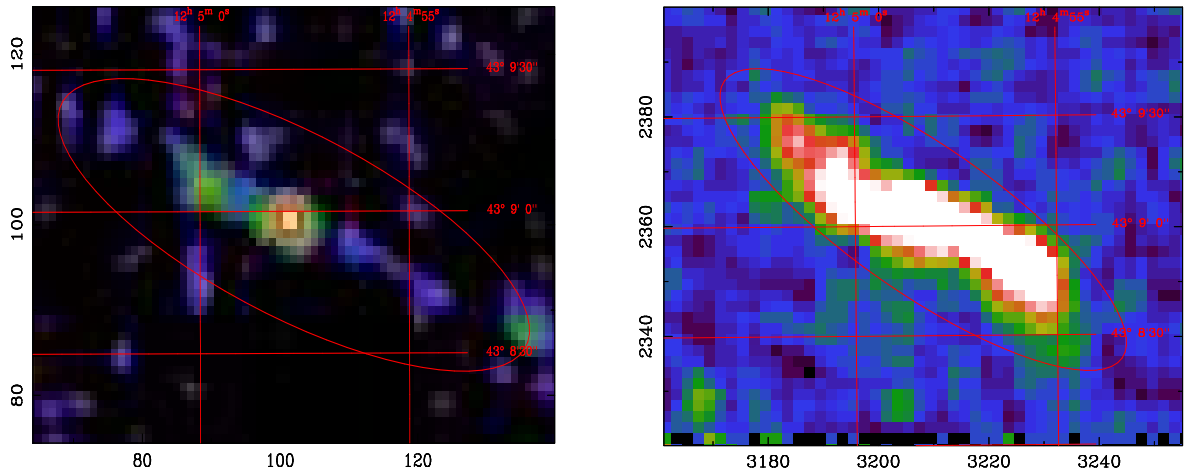

\centerline{\psfig{figure=new_large_lagaxy_FIRST.ps,angle=-90,height=6.15cm,width=7.25cm}
\hskip 1cm
            \psfig{figure=new_large_galaxy_NUV.ps,angle=-90,height=6.15cm,width=7.25cm}}
\caption{Left-hand panel (Fig. 4a) shows the VLA/FIRST image of UGC~7069. The ellipse has same dimensions as that in Fig. 2a.  Right-hand panel (Fig. 4b) displays the GALEX NUV (1750 - 2750~\AA) band image of UGC~7069. The major- and minor-axis diameters of the ellipse are 134".0 and 42".0, respectively, which indicate that the size of this galaxy is larger in the FUV band than that of the optical $g$ band.}
\end{figure*}

FITS images of UGC~7069 in the $J$, $H$ and $K_{s}$ bands were retrieved from the 2MASS database. Astrometric and photometric analyses were performed with these images using the $LEXTRCT$ (Tennant 2006). Integrated $J$, $H$ and $K_{s}$ band brightnesses of UGC~7069 are 12.91$\pm$0.03 (11.0$\pm$0.34 mJy), 12.29$\pm$0.04  (12.2$\pm$0.49 mJy) and 11.89$\pm$0.06  (11.3$\pm$0.61 mJy), respectively. These results are consistent with the catalogue values (Skrutskie et al. 2006).
The corresponding absolute magnitudes in the $J$, $H$ and $K_{s}$ bands and the $K$ band luminosity (L$_{K}$)
are -23.86, -24.48 and -24.88 mag 
and 8.34$\pm$0.45$\times$10$^{44}$~ergs~s$^{-1}$, respectively. 
L$_{K}$ is approximately one order of magnitude
brighter than that of the Cartwheel galaxy
(L$_{K}$=1.26$\pm$0.16$\times$10$^{44}$~ergs~s$^{-1}$).

UGC~7069 was detected in both the radio surveys with VLA/Faint Images of the radio Sky at the Twenty-centimeters (FIRST) and  the NRAO VLA Sky Survey (NVSS). The VLA/FIRST peak flux density of the nuclear region is 1.29$\pm$0.13 mJy (White et al. 1997) and the integrated flux density measured with the 45".0 beam size of NVSS is 6.2$\pm$0.5 (Condon et al. 1998). Thus, we do not know the integrated flux density of the whole galaxy. However, using the NVSS value we can compute the approximate far-infrared (FIR)
luminosity (L$_{FIR}$) of UGC~7069. Based on equations (14) and (15) of Condon, Cotton \& Broderick (2002),  we find that the value of L$_{FIR}$ will be 2.97$\pm$0.2$\times$10$^{44}$~ergs~s$^{-1}$.  We know for sure that this is not the total FIR luminosity of UGC~7069, because of incompleteness in the radio measurements. Unfortunately, UGC~7069 was not covered in the $IRAS$ survey. However, using this value of L$_{FIR}$, the computed value of the SFR is 13.4 \msun  yr$^{-1}$ (Kennicutt 1998). This is  consistent with the SFR of other ring galaxies, including Cartwheel (Marston \& Appleton 1995; Dopita et al. 2002; Mayya et al. 2005). VLA/FIRST image of UGC~7069 is shown in Fig.~4a with the $D_{23}$ isophote, similar to that of Fig.~2a.
\begin{figure*}
\centerline{\psfig{figure=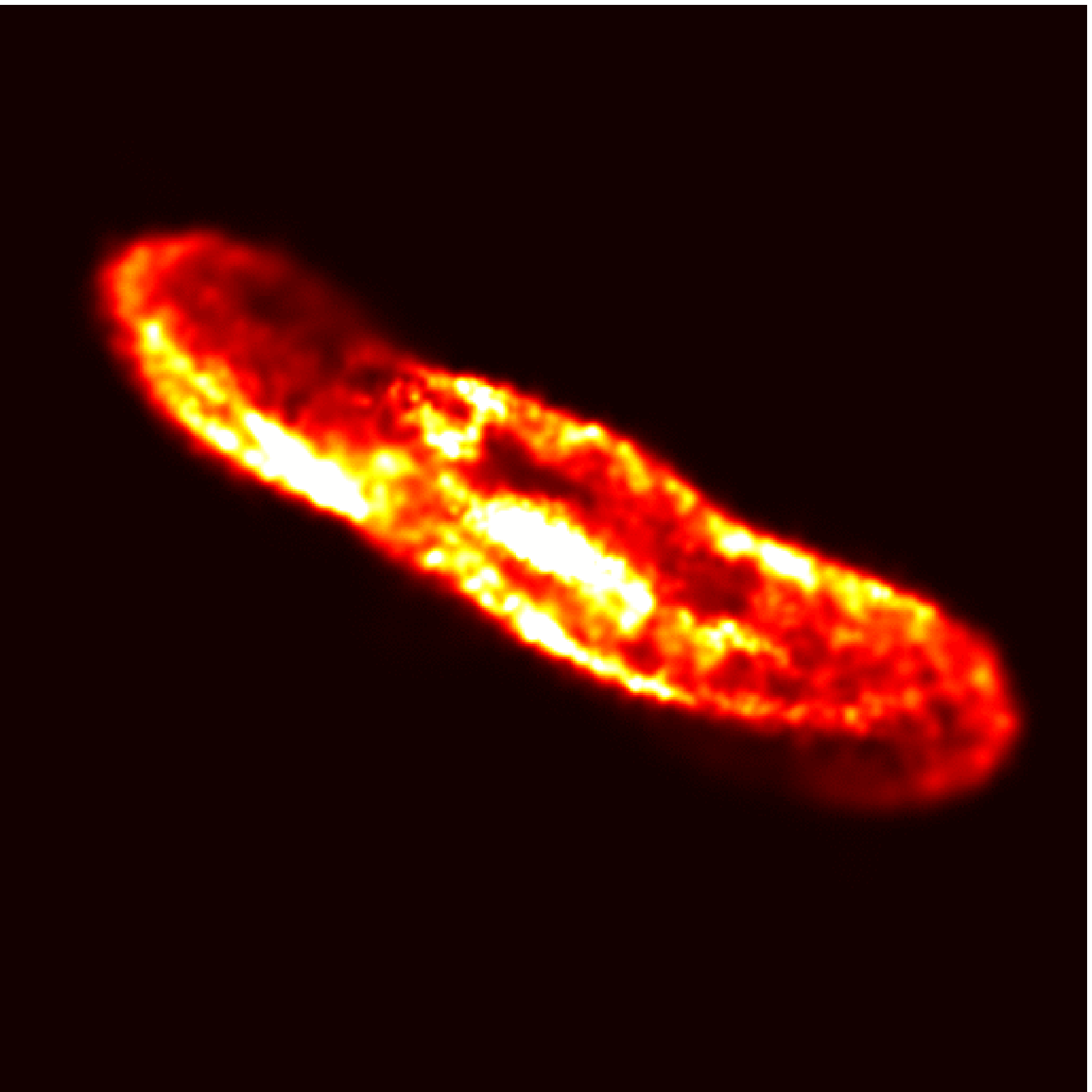,angle=0,width=6.2cm}
\hskip 1cm
          \psfig{figure=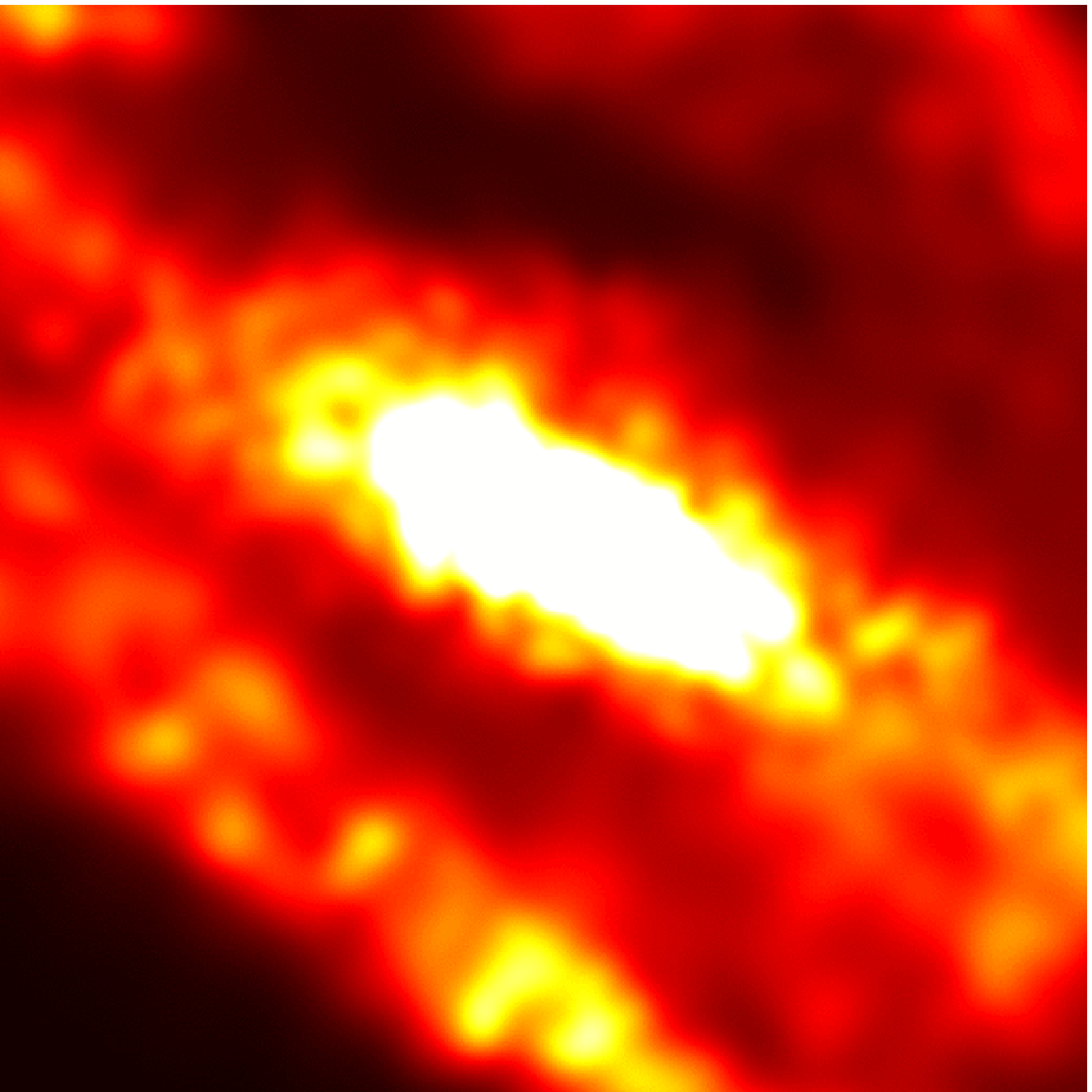,angle=0,width=6.2cm}}
\caption{Left-hand panel (Fig.~5a): density map of stars, in the flyby simulation, at $t=430$ Myr. The galaxy has been rotated by 85$^\circ{}$. The frame measures  125 kpc per edge. The density ranges from 0 to 18 \msun  pc$^{-2}$ in linear scale. 
Right-hand panel (Fig.~5b):  zoom of Fig.~5a in the nuclear region.
The frame measures 20 kpc per edge.
The density ranges from 0  to 33 \msun  pc$^{-2}$  in linear scale.}
\end{figure*}

GALEX (Martin et al. 2005) images are centred at 2312~\AA{} for the near-UV (NUV, 1750 - 2750~\AA) and at 1529~\AA{} for the far-UV (FUV, 1350 - 1750 \AA) bands. These images of UGC~7069 were retrieved from the $GALEX$ database and were analyzed using the $LEXTRCT$ (Tennant 2006). Fig.~4b displays the GALEX/NUV-band image of UGC~7069. The ellipse  with the major- and minor-axis diameters as 134".0 (133 kpc at the source frame) and 42".0 (42 kpc at the source frame), respectively, are shown in this figure. Derived luminosities are 8.13$\pm$0.4 and 7.38$\pm$0.4$\times$10$^{43}$~ergs~s$^{-1}$ in the NUV and FUV bands, respectively. Thus, the UV luminosity of UGC~7069 will be 1.55$\pm$0.06$\times$10$^{44}$~ergs~s$^{-1}$, which is comparable to those of 2MASS and FIR luminosities. The NUV and FUV flux densities were used to compute the UV spectral index ($\beta_{26}$) using equation (2) of Calzetti et al. (2005). The derived value of $\beta_{26}$ is -0.92$\pm$0.15, which corresponds to strong blue colour (Calzetti et al. 2005) and is suggestive of intense star formation.
Unfortunately, no X-ray data are available for this galaxy. However, the expected X-ray luminosity in the $2 - 10$ keV band can be computed using the value of the  SFR  ($\sim{}$13.4 \msun  yr$^{-1}$) in  equation (22) of Grimm, Gilfanov \& Sunyaev (2003).  Estimated value of the $2 - 10$ keV band X-ray luminosity is $\sim{}2.0\times10^{41}$~ergs~s$^{-1}$, which will be $\sim3.3\times10^{41}$~ergs~s$^{-1}$  in the 0.2-10~keV band, assuming a power law of photon index 1.7 and the Galactic value of N$_{H}$ ($1.33\times10^{20}$~cm$^{-2}$).
This suggests that UGC~7069 is a potential host of ultraluminous X-ray sources (ULXs) and some of them may be intermediate-mass black hole systems (Mapelli et al. 2008a). Thus, it would be interesting to obtain {\it Chandra} and/or {\it XMM-Newton} data of this galaxy.


\section{Results from Simulations}

We simulate a $N$-body/SPH model of UGC~7069  to reproduce its main peculiarities, that is i) the huge extension of the ring, ii) the warped edges, and iii) the double nucleus.
Details of simulations are described in Mapelli et al. (2008a, 2008b). Here we mention only the input parameters and present the corresponding results. 
We simulate the collision between a disc galaxy (target) and a smaller companion (intruder). The target galaxy is composed of a Navarro, Frenk \& White (1996, NFW) dark matter halo, a stellar and a gaseous Hernquist disc (Hernquist 1993). Its properties are the same as described in run B3 of Mapelli et al. (2008a): the virial mass, the virial radius and the concentration of the NFW halo are  $M_{200}=4.9\times{}10^{11}\,{}M_\odot{}$, $R_{200}=140$ kpc and $c=12$, respectively; the stellar disc mass, scalelength and scaleheight are $M_{d}=4.8\times{}10^{10}\,{}M_\odot{}$, $R_{d}=6.6$ kpc and $z_0=0.2\,{}R_d$, respectively; the analogous quantities for the gaseous disc are $M_g=3.2\times{}10^{10}M_\odot{}$, $R_g=R_d$ and $z_g=0.057\,{}R_g$. 
The intruder is composed of a NFW halo (with virial mass $M_{200}=3.2\times{}10^{11}\,{}M_\odot{}$, virial radius $R_{200}=30$ kpc and  $c=12$) and of a gaseous Hernquist disc (with mass $M_g=2\times{}10^{9}M_\odot{}$, scalelength $R_g=1.32$ kpc and scaleheight $z_g=0.1\,{}R_g$).
The target as well as the intruder are bulgeless. The gas of the target and of the intruder is allowed to form stars, according to the recipe described by Katz (1992), with an efficiency $c_\ast{}=0.1$.
In all the simulations we adopt a null impact parameter, as the two sides of UGC~7069 are symmetric around the nucleus, and the ring is probably circular, apart from the warps, once accounted for the inclination with respect to the observer.

We focus on two different scenarios: i) a merger between intruder and target; and ii) a flyby interaction in which the angle between the relative velocity of the two galaxies ({\bf v$_{rel}$}) and the disc axis of the target is large ($30^\circ{}-60^\circ{}$). The first scenario has been considered to check whether the two nuclei of UGC~7069 are due to an on-going merger. In order to have a merger,  {\bf v$_{rel}$} has to be smaller than the escape velocity. Thus, we assume that the modulus of {\bf v$_{rel}$} is $v_{rel}=300$ km s$^{-1}$.  
The largest ring that we obtain in our merger simulations reaches a diameter of $\sim{}70$ kpc. Thus, the perturbation is not strong enough to produce a ring as large as the one observed in UGC~7069, although this discrepancy between simulations and observations is not sufficient to discard the merger scenario.
Furthermore, the ring is not warped. In addition, the nucleus does not appear double at this stage, as the intruder completely merged before the ring-galaxy phase. We stress that the minimum length that we can resolve with our simulations is $\sim{}100$ pc (i.e. the softening length): the nucleus might be double below this scale. However, the two observed nuclei are separated by $\sim{}2.9$'' (i.e. $\sim{}3$ kpc) and should be resolved by the simulation. 
Of course, these results depend on the initial conditions. Smaller values of  $v_{rel}$ produce even smaller rings, whereas for larger values of  $v_{rel}$ the merger does not occur during the first approach. We could hypothesize that the ring is produced during the first approach and the merger occurs later. However, this scenario requires that the second approach and the merger occur within $\lesssim{}500$ Myr after the formation of the ring; otherwise the ring will disappear. Even if we suppose that this scenario is able to produce the double nucleus, it seems quite fine-tuning.

As a second scenario, we simulate flyby interactions between the target and the intruder. In this case, we adopt $v_{rel}=900$ km s$^{-1}$. We make various simulations with inclination angle ranging from 30$^\circ{}$ to 60$^\circ{}$. A large inclination is required to produce warps. In this case, we obtain rings with a diameter as large as  $\sim{}120$ kpc. This is an important result, as we demonstrate that galaxy interactions can produce huge rings. In Fig.~5a we show the stellar density in one of our simulations (with inclination angle between intruder and target equal to 45$^\circ{}$), at  $\sim{}$430 Myr after the galaxy collision. The diameter is $\sim{}110$ kpc. The SFR derived from this simulation is $\sim{}9$ M$_\odot{}$ yr$^{-1}$, approximately in agreement with the value estimated from observations.
The ring appears slightly warped at its edges. Zooming in the central region (Fig.~5b), the nucleus appears strongly perturbed and knotty, even if it is not properly double-peaked. Thus, the nucleus of a ring galaxy can be strongly perturbed by the interaction. This suggests that even the nucleus of UGC~7069 can appear double-peaked as a consequence of the galaxy collision. Furthermore, one of the two observed peaks might also be associated with a strong HII region or with the secondary ring, whereas the other one is the `real' nucleus. Finally, even strong absorption from dust in the central region might produce the double-peak appearance.

\section{Discussion and Conclusions}
In this Letter we show that UGC~7069 is a huge collisional ring galaxy, with a diameter of $\sim{}115$ kpc. This galaxy has other interesting features: i) the ring is warped at its edges; ii) the nucleus appears double-peaked and it is probably a LINER. We stress that UGC~7069 is the first ring galaxy to be classified as a LINER. However, the nature of the double nucleus is unclear. The two peaks might be two distinct nuclei, or a nucleus and a bright HII region, or a single nucleus with strong absorption, or a single nucleus heavily perturbed by the galaxy interaction. Thus, further investigations are essential.
The VLA/FIRST and the NVSS data indicate quite intense and extended radio emission. The UV luminosity is high and the UV spectral index extremely steep ($\beta_{26}=-0.92\pm0.15$), indicating strong blue colour. This suggests a high SFR for this galaxy. In an indirect way, we estimated the SFR to be $\sim{}13\,{}{\rm M}_\odot{}$ yr$^{-1}$. 

The simulations indicate that rings with a diameter of  $\sim{}120$ kpc can be produced by flyby interactions between galaxies. 
Interactions where the inclination between the intruder and the target is $\sim{}45^\circ{}$ can also produce warped rings. Furthermore, the nucleus of the simulated galaxy appears strongly perturbed by the interaction. The simulated SFR is also quite high ($\sim{}9\,{}$M$_\odot{}$ yr$^{-1}$).
If  UGC~7069 is a collisional ring galaxy, then another question to be addressed is the nature of the intruder. If the intruder galaxy has not merged with the target (as simulations suggest), if its relative velocity with respect to UGC~7069 is $\lesssim{}1000$ km s$^{-1}$ and if the interaction occurred $\lesssim{}500$ Myr ago, its current projected distance from UGC~7069 should be $d_{proj}\lesssim{}500$ kpc. In the SDSS image we found at least five galaxies with a similar $d_{proj}$ from UGC~7069: SDSS J120433.94+430611.1 ($d_{proj}\sim{}314$ kpc and unknown redshift), SDSS J120432.24+430307.2 ($d_{proj}\sim{}453$ kpc and $z=0.054$), SDSS J120515.56+431008.4 (an elliptical galaxy with $d_{proj}\sim{}212$ kpc  and $z=0.053$, probably too massive to be the intruder), SDSS J120517.29+430534.8 ($d_{proj}\sim{}298$ kpc and $z=0.055$), SDSS J120523.31+431107.5 ($d_{proj}\sim{}313$ kpc and $z=0.051$). Further investigations are required to find whether or not the intruder is one of these galaxies.

In conclusion, UGC~7069 appears to be a very interesting object. The existence of such huge ring is also an indirect proof that ring galaxies might evolve into giant low surface brightness galaxies (Mapelli et al. 2008b), with very large and flat discs (radius $\sim{}100-150$ kpc). Thus, the properties of UGC~7069 deserve further observations and investigations. 
For example, it would be important to carry out X-ray observations of this galaxy, in order to determine the nature of the nuclear region, to detect ULXs and to determine the global X-ray properties of this galaxy.

\section*{Acknowledgments}
We thank Ben Moore, Lea Giordano and Emanuele Ripamonti for useful discussions  and we acknowledge the anonymous referee for his helpful comments.
This research has used data from the SDSS, the 2MASS, the GALEX archive,  the VLA/FIRST and NVSS surveys.
The simulations have been carried out using the cluster {\it zbox2} at the University of Z\"urich.
MM acknowledges support from the Swiss
National Science Foundation.

\end{document}